\documentclass[10pt,aps,prd,noshowpacs,eqsecnum,nofootinbib,twocolumn,floatfix]{revtex4-1}
\usepackage[utf8]{inputenc}
\usepackage{amsmath,amsthm,amssymb}
\usepackage{mathrsfs,bbm}

\usepackage{nicefrac} 
\usepackage{latexsym}
\usepackage{array}
\usepackage{comment}

\usepackage{color}
\usepackage{multirow}
\usepackage{caption}
\DeclareCaptionJustification{justified}{\leftskip=0pt \rightskip=0pt \parfillskip=0pt plus 1fil}
\begin{document}


\title{CKM mixings from mass matrices with five texture zeros
}

\author{Yithsbey Giraldo}
\email[E-mail: ]{yithsbey@gmail.com}
\author{Eduardo Rojas}
\email[E-mail: ]{eduro4000@gmail.com}
\affiliation{Departamento de F\'\i sica, Universidad de Nari\~no, A.A. 1175, San 
Juan de Pasto, Colombia}
   
\date{\today}

\begin{abstract}
In this work we carry out  an exhaustive study to find quark mass matrices in the Standard Model~(SM), with the maximum number of texture zeros consistent with the experimental data. We found four viable configurations  of five texture zeros that adjust the quark masses, the mixing angles and the CP violation phase, with deviations below $1\sigma$ level respect to the current SM best fit values.
One of the most important aspects of this work is an economic procedure to find the texture zeros: we resort to the weak basis transformation method, which, as we will show,   exhaustively search  every  possible configuration. We report various leading order relations between the mixing angles and the quark masses for each case.
\end{abstract}

\pacs{12.15.Ff}

\maketitle

\section{Introduction}
In the Standard Model~(SM), the quark mass matrices come from  the interaction  between the 
Higgs boson and the SM fermions. After the spontaneous breaking of the SM gauge symmetry we obtain
\begin{equation}
 -{\cal L}_M=\bar u_RM_u u_L+\bar d_RM_dd_L+h.c.,
\end{equation}
where  $ M_u $ and $ M_d$  are arbitrary,   $ 3\times3 $ quark 
mass matrices containing thirty-six~(36) real parameters,
which cannot be fully determined from the ten~(10) physical 
observables that they must account for: six~(6) quark masses, three~(3) flavor 
mixing angles, and one~(1)  charge-parity~(CP) violating phase. However, in 
models like the SM~(or its extensions) where the right fields are singlets under 
the gauge group, it is always possible to choose a suitable basis for the right 
quarks, such that by using the {\it the polar decomposition theorem} of the matrix 
algebra, the mass matrices of type ``up'' and ``down'' became 
hermitian~\cite{Fritzsch:1999ee,Xing:2015sva,Gupta:2013yha,Ludl:2015lta,Giraldo:2011ya,Fusaoka:1998vc}.
\begin{equation}
\label{1.2x}
 M_u^\dag=M_u,\quad  \textrm{and}\quad M_d^\dag=M_d.
\end{equation}
Additionally, for Hermitian quark mass matrices, you can 
make a unitary transformation acting simultaneously on the up-type and 
down-type  quark mass matrices, leaving the gauge currents invariant, and  the mass 
matrices transform to new equivalent Hermitian matrices
\begin{equation}
\label{1.2}
 M_u\rightarrow M'_u=U^\dag M_uU,\quad M_d\rightarrow M'_d=U^\dag M_d U,
\end{equation}
where $U$ is an arbitrary unitary matrix that preserves the hermiticity of the 
mass matrices and leaving  the physical quantities invariant, in particular,  the
Cabibbo-Kobayashi-Maskawa~(CKM) mixing matrix.
This common unitary transformation applied to $M_u$ and $M_d$, in Eq.~\eqref {1.2},
is known as a ``Weak Basis'' (WB) 
transformation~\cite{Branco:1999nb,Fritzsch:1999ee,b_1,b4x,Verma:2013qta}.
As it was shown in~\cite{Giraldo:2011ya,Giraldo:2015ffa}, 
for a given set of quark masses, mixing angles and the CP-violating phase, all the 
mass matrices consistent with these experimental values are unitarily 
equivalent.  This result can be used to calculate the maximum number of texture 
zeros, since it guarantees that by using  WB transformations it is 
possible to reach all
physical and non-physical zeros consistent with the 
data~\cite{Branco:1999nb,Giraldo:2011ya}. 
Through a  WB transformation, it is possible to rewrite  the quark mass matrices  as follows~\cite{Branco:1999nb,Giraldo:2011ya,Giraldo:2015dta,Giraldo:2015ffa}:
\begin{subequations}
\label{2.8}
\begin{equation}
\label{3}
\begin{split}
 M_u&=D_u=\begin{pmatrix}
          \lambda_{1u}&0&0\\
0&\lambda_{2u}&0\\
0&0&\lambda_{3u}
         \end{pmatrix},\\
%
 M_d&=VD_dV^\dag,
\end{split}
\end{equation}
or
\begin{equation}
\label{4}
\begin{split}
 M_u&=V^\dag D_uV,\\
%
 M_d&=D_d=\begin{pmatrix}
          \lambda_{1d}&0&0\\
0&\lambda_{2d}&0\\
0&0&\lambda_{3d}
         \end{pmatrix},
\end{split}
\end{equation}
\end{subequations}
where $V=U_u^{\dagger} U_d$ is the CKM mixing matrix,
$U_u$  and $U_d$ are the diagonalization matrices for the mass matrices  $M_u$ and $M_d$, respectively. 
The parameters $\lambda_{iq}$ ($i=1,2,3$) are the quark mass matrix eigenvalues 
for up-type~($q=u$) and down-type~($q=d$) quarks, which are related to the quark masses 
\begin{equation}
\label{2.9}
 \begin{split}
|\lambda_{1u}|=m_u, |\lambda_{2u}|=m_c, |\lambda_{3u}|=m_t,
\\
|\lambda_{1d}|=m_d, |\lambda_{2d}|=m_s, |\lambda_{3d}|=m_b.
\end{split}
\end{equation}
So  $\lambda_ {iq}$ can be positive or negative and obey the hierarchy
\begin{equation}
\label{19d}
|\lambda_{1q}|\ll|\lambda_{2q}|\ll|\lambda_{3q}|.
\end{equation}
In the basis~\eqref{2.8}  can be easily verified that the mass matrices are consistent with 
the CKM mixing matrix $V$ and the quark masses,  and the 3 non-physical texture zeros can be 
effortlessly identified~\cite{Branco:1999nb}.
The hermiticity of the quark mass matrices $ M_u $ and $ M_d $ reduces the 
number of 
free parameters from 36 to 18, which, however, is still a large value compared 
to the number of observables. In order to reduce the number of free parameters, 
Weinberg and Fritzsch~\cite{Fritzsch:1979zq, b31, b32, b33} introduced 
texture-zeros into the mass matrices with a dual purpose, first of all, to 
obtain  self-consistent relationships between the quark masses and the flavor mixing parameters 
that can be experimentally verified~\cite{Fritzsch:1999ee,Fritzsch:2002ga}. On the other hand, the discrete~(or continuous) flavor symmetries hidden in 
such textures may finally provide clues on the origin of the energy scales in 
the quark sector of the SM as residual symmetries of a more fundamental 
symmetry at high energies.
Hermitian quark mass matrices with six texture zeros were introduced in what is currently known as the Fritzsch 
type~\cite{b33,Fritzsch:2002ga},
where the mass matrices, $M_u$, and $M_d$, have the same texture (``up-down'' parallel) each with three zeros. This type of ansatz was ruled out due to the  
large value of the mass of the top quark, since that for this case the CKM element 
$|V_{cb}|$ is in tension with  the
experimental data~\cite{Gupta:2013yha,b44,Fritzsch:2002ga}. 
Furthermore, for reasonable values of the current quark masses $m_u$ and $m_c$, 
the expected magnitude for $|V_{ub}/ V_ {cb} |=\sqrt {m_u/m_c}\approx 0.05$~\cite{Xing:2011aa} 
is too small  in comparison  with the 
experimental value ($ |V_ {ub}/V_{cb}| _ {\text{exp.}}\approx 0.09 
$~\cite{Tanabashi:2018oca,UTfit,CKMfitter}). 
In this sense, one of the difficulties of working with texture zeros is keeping the predictions for $ V_{us} $ and $ V_{cd}$ right, and simultaneously reproducing the ratios
$V_{ub}/V_{cb}$ and $V_{td}/V_{ts}$,  i.e.,
\begin{equation}
\begin{split}
 \left|\frac{V_{ub}}{V_{cb}}\right|_{\text{exp.}}&=0.0861\pm0.0027,\\
 \left|\frac{V_{td}}{V_{ts}}\right|_{\text{exp.}}&=0.2107\pm0.0044.
\end{split}
\label{eq1}
\end{equation}
The original literature on  five-zero textures has been widely studied, 
but these initial ansatzes are not currently favored by 
experimental data~\cite{Fakay:2014rea,Sharma:2014tea,Sharma:2015gfa,Gupta:2013yha,
Randhawa:1999hi,Branco:1999nb,Desai:2000bu,Mahajan:2009wd}.
Recent studies show that other five-zero textures are viable, some analytical 
and numerical examples were reported in~\cite{Verma:2017ppl,Giraldo:2011ya,Ludl:2015lta,Ponce:2013nsa,Ponce:2011qp}, 
these textures  reproduce the quark masses and the CKM mixing matrix  with deviations respect to the 
experimental values below 1$\sigma$ level.
There are several approaches to obtain the texture zeros, in some cases, the analytic 
approximations  take advantage of the strong hierarchy in quark 
masses and mixing angles to motivate a certain texture~\cite{Fritzsch:1986sn,Fritzsch:2002ga}, 
alternatively, some  techniques prefer to assume a texture for the quark mass matrices  to make physical predictions~\cite{Ponce:2011qp,Verma:2017ppl,Gupta:2013yha,Hernandez:2014zsa}. A very elegant way is to apply WB 
transformations in order to get texture zeros in 
the mass matrices~\cite{Branco:1999nb,Giraldo:2011ya}, 
our work points in this  direction and it can be considered as a continuation of the work  presented
by one of us in~\cite{Giraldo:2011ya}.
This work is organized as follows:
In Section~\ref {sIII} we classify all possible ways to put three texture-zeros in the  ``up'' or ``down'' quark mass matrices. This analysis is important since from these textures we can obtain five texture zeros for the mass matrices by using the WB transformation method.
We will carry out a first analytical study for  five-zero textures in Section~\ref{sIV}, and the  conclusions are summarized  in Section~\ref{sV}.
\section{Five-zero textures}
\label{sIII}
In order to maintain the determinant different from zero the mass matrix for up quarks (or down quarks)  has at most three texture-zeros~\footnote{More than three texture zeros implies that at least one quark mass is equal to zero or  two of the quark masses must be equal~\cite{Giraldo:2011ya,Ponce:2011qp}.}. Also, we only have two types of realistic patterns depending on how the three texture zeros are distributed in the inputs of the mass matrix. In the first case we have a matrix with two texture zeros on the diagonal, and in the other case the matrix only contains a texture-zero on the diagonal, as it is pointed out  in each column
of Table~\ref{ta1}; where it is shown that by doing WB transformations with the  permutation matrices $p_i$, with $i=1,\cdots,6$, we obtain all possible viable cases for each pattern. Table~\ref{ta1} summarizes all the viable three-zero textures (via permutations) for  the up and down quark mass matrices.
 Without loss of generality, as we will see later,  we can write these patterns without including phases.
\begin{table}[htb]
{\footnotesize
\begin{tabular}{|p{2.09cm}||p{3.17cm}||p{3.02cm}|}
\hline
\hline
\textbf{Permutation matrices}&\textbf{Pattern with two zeros on the diagonal}  
$(p_i\:M_q\:p_i^T)$ & \textbf{Pattern with one zero on the diagonal} $(p_i\:M_q\:	
p_i^T)$\\[1mm]
\hline
& &\\[-2mm]
$p_1=\begin{pmatrix}
      1&&\\
&1&\\
&&1
     \end{pmatrix}
$&$\begin{pmatrix}
      0& |\xi_q|& 0\\
|\xi_q|& 0& |\beta_q|\\
0& |\beta_q|& \alpha_q
     \end{pmatrix}$&
$\begin{pmatrix}
      0& |\xi_q|& 0\\
|\xi_q|& \gamma_q& 0\\
0& 0& \alpha_q
     \end{pmatrix}$\\[5mm]
\hline
& &\\[-2mm]
$p_2=\begin{pmatrix}
      1&&\\
&&1\\
&1&
     \end{pmatrix}
$&$\begin{pmatrix}
      0&0 & |\xi_q|\\
0& \alpha_q& |\beta_q|\\
|\xi_q|& |\beta_q|& 0
     \end{pmatrix}$&
$\begin{pmatrix}
 0& 0& |\xi_q|\\
0& \alpha_q& 0\\
|\xi_q|& 0& \gamma_q
     \end{pmatrix}$\\[5mm]
\hline
& &\\[-2mm]
$p_3=\begin{pmatrix}
     &&1 \\
&1& \\
1&& 
     \end{pmatrix}
$&$\begin{pmatrix}
      \alpha_q& |\beta_q|& 0\\
|\beta_q|& 0&|\xi_q| \\
0& |\xi_q|& 0
     \end{pmatrix}$&
$\begin{pmatrix}
      \alpha_q& 0& 0\\
0& \gamma_q& |\xi_q|\\
0& |\xi_q|& 0
     \end{pmatrix}$\\[5mm]
\hline
& &\\[-2mm]
$p_4=\begin{pmatrix}
     &1& \\
1&& \\
 &&1
     \end{pmatrix}
$&$\begin{pmatrix}
      0& |\xi_q|&|\beta_q|\\
|\xi_q|& 0& 0\\
|\beta_q|& 0& \alpha_q
     \end{pmatrix}$&
$\begin{pmatrix}
      |\gamma_q|& |\xi_q|& 0\\
|\xi_q|& 0& 0\\
0& 0& \alpha_q
     \end{pmatrix}$\\[5mm]
\hline
& &\\[-2mm]
$p_5=\begin{pmatrix}
      &&1 \\
1&& \\
&1&
     \end{pmatrix}
$&$\begin{pmatrix}
      \alpha_q& 0& |\beta_q|\\
0& 0& |\xi_q|\\
|\beta_q|& |\xi_q|& 0
     \end{pmatrix}$&
$\begin{pmatrix}
      \alpha_q& 0& 0\\
0& 0& |\xi_q|\\
0& |\xi_q|& \gamma_q
     \end{pmatrix}$\\[5mm]
\hline
& &\\[-2mm]
$p_6=\begin{pmatrix}
     &1& \\
&&1\\
1&& 
     \end{pmatrix}
$&$\begin{pmatrix}
      0&  |\beta_q|&|\xi_q|\\
 |\beta_q|&\alpha_q&0 \\
|\xi_q|& 0& 0
     \end{pmatrix}$&
$\begin{pmatrix}
      \gamma_q& 0& |\xi_q|\\
0& \alpha_q& 0\\
|\xi_q|& 0& 0
     \end{pmatrix}$\\[5mm]
\hline\hline
\end{tabular}
}
\caption{Mass matrix patterns with three texture-zeros. 
We are considering two cases, depending on the number of zeros in the diagonal (one or two texture zeros). It is not necessary to include phases.}
\label{ta1}
\end{table}
%
An equivalence transformation through a permutation is a type of WB transformation, 
indeed, this fact allows us to find equivalent textures through
 permutations, for example
\begin{equation}
 \begin{split}
  M'_u&=
\begin{pmatrix}
 0 & \times & 0 \\
 \times & \times & \times\\
 0 & \times & \times \\
\end{pmatrix}=p_2\cdot \begin{pmatrix}
 0 & 0 & \times \\
 0 & \times & \times \\
 \times & \times & \times \\
\end{pmatrix}\cdot p_2^T,\\
M'_d&=
\begin{pmatrix}
 0 & 0 & \times \\
 0 & \times & \times \\
 \times & \times &0 \\
\end{pmatrix}=p_2\cdot  \begin{pmatrix}
 0 & \times & 0 \\
 \times & 0 & \times \\
 0 & \times & \times \\
\end{pmatrix} \cdot p_2^T,
 \end{split}
\end{equation}
where ``$\times$'' stands for the non-zero entries.
It is important to mention that the permutations do not change the number of zeros on the diagonal.

We will work with five-zero textures for the quark mass matrices. 
 Six-zero textures have already been ruled out~\cite{Giraldo:2011ya,Kaundal:2019njx,Randhawa:1999hi,Mahajan:2009wd}.
\subsection{Texture-zero patterns}  
\label{sIIIC}
The patterns shown in Table~\ref {ta1} can be analytically diagonalized.
To accomplish this, we consider the most general case of a symmetric mass matrix with two texture zeros
\begin{equation}
\label{30}
 M_q=\begin{pmatrix}
      0&|\xi_{q}|&0\\
|\xi_{q}|&\gamma_{q}&|\beta_{q}|\\
0&|\beta_{q}|&\alpha_{q}
     \end{pmatrix},
\end{equation}
where the phases of the off-diagonal parameters  can be absorbed~(or included) in only one of the mass matrices (the down-type or the up-type) through a WB transformation. $\gamma_{q}$ and $\alpha_{q}$
are real numbers due to the hemiticity of $M_q$. According to the Table~\ref{ta1}, the pattern with two zeros on the diagonal
is achieved by making $\gamma_q = 0$, and to obtain the pattern with a zero on the diagonal we set $ |\beta_q| = 0 $.
The mass matrix $M_q$ can be diagonalized using the
transformation
\begin{equation}
\label{31x}
U_q^\dag M_qU_q=D_q=
\begin{pmatrix}
          \lambda_{1q}&&\\
&\lambda_{2q}&\\
&&\lambda_{3q}
         \end{pmatrix},
\end{equation}
where the $\lambda_ {iq}$~$ (i = 1,2,3)$ are defined in~\eqref{2.9}. Note that $\gamma_{q}$, $|\beta_{q}|$ and $|\xi_{q}|$ can be expressed in terms of $\alpha_q$ and the $\lambda_{iq}$'s. 
By using the invariants under a basis transformation, $\text {tr} M_q$, $\textrm{tr} M_q^2$ and $\det M_q$, it follows that
\begin{subequations}
\label{e3.4}
\begin{align}
\label{3.18y}
 \gamma_q&=\lambda_{1q}+\lambda_{2q}+\lambda_{3q}-\alpha_q,\\
\label{34ay}
|\beta_q|&=\sqrt{\frac{(\alpha_q-\lambda_{1q})(\alpha_q-\lambda_{2q})(\lambda_{
3q}-\alpha_q)}{\alpha_q}},\\
\label{35ay}
|\xi_q|&=\sqrt{\frac{-\lambda_{1q}\lambda_{2q}\lambda_{3q}}{\alpha_q}}.
\end{align}
\end{subequations}
According to~\cite{Branco:1999nb,Giraldo:2011ya,Giraldo:2018mqi} and the relation~\eqref{35ay}~(which is real), $\alpha_q>0$; and from~\eqref{34ay}, it must be found in one of the following intervals:
{\small
\begin{subequations}
\label{e3.5s}   
\begin{equation}
\label{e3.5}
\text{If}\   \lambda_{1q}<0, \lambda_{2q}>0 \ \text{and}\  \lambda_{3q}>0 \implies  |\lambda_{2q}|\le\alpha_q\le|\lambda_{3q}|
\end{equation}
\begin{equation}
\label{e3.6}
\text{If}\  \lambda_{1q}>0, \lambda_{2q}<0 \ \text{and}\   \lambda_{3q}>0 \implies
 |\lambda_{1q}|\le\alpha_q\le|\lambda_{3q}|.
\end{equation}
\begin{equation}
\label{e3.7}
\text{If}\  \lambda_{1q}>0, \lambda_{2q}>0 \ \text{and}\   \lambda_{3q}<0 \implies
 |\lambda_{1q}|\le\alpha_q\le|\lambda_{2q}|.
\end{equation}
\end{subequations}}
In the previous analysis, the~(\ref{19d}) hierarchy was taken into account, and we only considered a negative eigenvalue according to the justification given in papers~\cite{Giraldo:2011ya,Giraldo:2018mqi}~\footnote{

The WB transformations allow us to use the
basis~\eqref{3} (or the basis in~\eqref{4}) as the ``starting point'' matrices to generate any viable representation of quark mass matrices~\cite{Giraldo:2011ya,Giraldo:2015ffa}. If there are
texture zeros in mass matrices these can be found by a WB transformation.
Texture zeros on the diagonal of the mass matrices imply that at least one of
the proper values must be negative~\cite{Branco:1999nb,b4,b4x,Giraldo:2011ya}.
}.

The exact analytical matrix~$U_q$, which diagonalizes the mass matrix~\eqref{30}, 
is given by~\cite{b4, b4x,Giraldo:2011ya}
\begin{widetext}
\begin{equation} 
\label{32x}
{
 U_q=\begin{pmatrix}
     e^{i\theta_1} 
\frac{|\lambda_{3q}|}{\lambda_{3q}}\sqrt{\frac{\lambda_{2q}\lambda_{3q}
(\alpha_q-\lambda_{1q})}{\alpha_q(\lambda_{2q}-\lambda_{1q})(\lambda_{3q}
-\lambda_{1q})}}&e^{i\theta_2}\frac{|\lambda_{2q}|}{\lambda_{2q}}
\sqrt{\frac{\lambda_{1q}\lambda_{3q}(\lambda_{2q}-\alpha_q)}{\alpha_q(\lambda_{
2q}-\lambda_{1q})(\lambda_{3q}-\lambda_{2q})}}&
\sqrt{\frac{\lambda_{1q}\lambda_{2q}(\alpha_q-\lambda_{3q})}{\alpha_q(\lambda_{
3q}-\lambda_{1q})(\lambda_{3q}-\lambda_{2q})}}\\
&&&\\[-2mm]
-e^{i\theta_1}\frac{|\lambda_{2q}|}{\lambda_{2q}}\sqrt{\frac{\lambda_{1q}
(\lambda_{1q}-\alpha_q)}{(\lambda_{2q}-\lambda_{1q})(\lambda_{3q}-\lambda_{1q})}
}&
e^{i\theta_2}\sqrt{\frac{\lambda_{2q}(\alpha_q-\lambda_{2q})}{(\lambda_{2q}
-\lambda_{1q})(\lambda_{3q}-\lambda_{2q})}}&
\frac{|\lambda_{3q}|}{\lambda_{3q}}\sqrt{\frac{\lambda_{3q}(\lambda_{3q}
-\alpha_q)}{(\lambda_{3q}-\lambda_{1q})(\lambda_{3q}-\lambda_{2q})}}\\
&&&\\[-2mm]
e^{i\theta_1}\frac{|\lambda_{2q}|}{\lambda_{
2q}} \sqrt{\frac{\lambda_{1q}(\alpha_q-\lambda_{2q})(\alpha_q-\lambda_{3q})}{
\alpha_q(\lambda_{2q}-\lambda_{1q})(\lambda_{3q}-\lambda_{1q})}}&
-e^{i\theta_2}\frac{|\lambda_{3q}|}{\lambda_{3q}}\sqrt{\frac{\lambda_{2q}
(\alpha_q-\lambda_{1q})(\lambda_{3q}-\alpha_q)}{\alpha_q(\lambda_{2q}-\lambda_{
1q})(\lambda_{3q}-\lambda_{2q})}}&
\sqrt{\frac{\lambda_{3q}(\alpha_q-\lambda_{1q})(\alpha_q-\lambda_{2q})}{
\alpha_q(\lambda_{3q}-\lambda_{1q})(\lambda_{3q}-\lambda_{2q})}}
     \end{pmatrix},}
\end{equation}
\end{widetext}
where we have included additional phases~(non-physical) to adjust the~CKM mixing matrix to the
usual convention~\eqref{3.2}, as shown in the reference~\cite{Giraldo:2015dta}.
It is not necessary to include a phase in the third column, as it can be absorbed by the remaining phases.

The diagonalization matrix~\eqref {32x} can be seen as the unitary matrix  of a  WB transformation on the initial mass representations~\eqref{2.8}. For the case~\eqref{3}:
\begin{subequations}
\label{2.9x}
\begin{align}
M_{u}^{\prime}&=U_{u}(D_u)U_{u}^\dag=\begin{pmatrix}
      0& |\xi_{u}|& 0\\
|\xi_{u}|& \gamma_u& |\beta_{u}|\\
0& |\beta_{u}|& \alpha_{u}
     \end{pmatrix},
\\
\label{91ax}
M_{d}^\prime&=U_{u}\,(V D_dV^\dag)\,U_{u}^\dag\,,
\end{align}
\end{subequations}
where Eq.~\eqref{31x} was considered.
As we have already mentioned, if we want a pattern of three zeros in the mass matrix $ M'_u $, with two zeros on the diagonal, that is, with $\gamma_u=0$,   it is necessary to make $\alpha_u=\lambda_{1u}+\lambda_ {2u}+\lambda_{3u}$ according to~\eqref{3.18y}. 
From~\eqref{e3.5s} this configuration is only possible for $\lambda_{1u}, \lambda_{3u}>0$ and  $\lambda_ {2u}<0$.
To find two additional texture zeros in the inputs of the mass matrix~\eqref{91ax},
we adjust the free parameters $\theta_1 $ and $\theta_2$ of the diagonalization matrix~\eqref{32x}.
On the other hand, if we want three zeros for the mass matrix $M'_u$,
but with a single zero on the diagonal, it is necessary to set $|\beta_u| = 0$. To achieve this we have three possibilities~(from Eq.~\eqref{34ay}): $\alpha_u = \lambda_{1u}$, or $\alpha_u =\lambda_ {2u}$, or $\alpha_u = \lambda_{3u}$.
In each of these cases, one of the remaining $\lambda_{iu}$'s must be negative, which gives a total of six different possibilities. A similar exercise can be carried out in the case~\eqref{4}.
\begin{subequations}
\label{2.10}
\begin{align}
\label{91axx}
M_{u}^\prime&=U_{d}\,(V^\dag D_uV)\,U_{d}^\dag\,,
\\
M_{d}^{\prime}&=U_{d}(D_d)U_{d}^\dag=\begin{pmatrix}
      0& |\xi_{d}|& 0\\
|\xi_{d}|& \gamma_d& |\beta_{d}|\\
0& |\beta_{d}|& \alpha_{d}
     \end{pmatrix}.
\end{align}
\end{subequations}
where we have used the relation~\eqref{31x} for the special case $q=d$. Table~\ref{t2} summarizes the numerical results of our study, in the next section we will see these results in more detail from an analytical point of view.
\begin{widetext}
\begin{center}
{\footnotesize
\begin{tabular}{c|c|c|c|c|p{6.2cm}|}
\hline
\hline
\multirow{4}{*}{\bf Case}&\multirow{4}{*}{\bf Five-zero textures}&&\multirow{4}{*}{{\bf Best fit values}~(MeV)}&\multirow{3}{*}{\bf Negative}&\centerline{\bf Pulls:}
\\[-3.5mm]
&&&&\multirow{3}{*}{\bf eigenvalues}&\multirow{3}{*}{ 
\begin{tabular}{lllll}
 Wolfenstein parameters:& $P_\lambda$ &$P_A$ & $P_{\rho}$&$P_{\eta}$ \\
 Up-type quark masses:&$P_{m_u}$ &$P_{m_c}$&$P_{m_t}$&$-$ \\
 Down-type quark masses:&$P_{m_d}$ & $P_{
m_s}$ &$ P_{m_b}$&$-$  \\
\end{tabular}}\\[4.0mm]
&&&&&
\\\hline
&&&&&\\[-2mm]
\multirow{10}{*}{\bf I}&\multirow{5}{*}{
$M_{Iu}=
\begin{pmatrix}
 0&0&\xi_u\\
0&\alpha_u&\beta_u\\
\xi_u^*&\beta_u^*&\gamma_u
\end{pmatrix}$
}&\multirow{6}{*}{\bf a.}
&\multirow{3}{*}{\begin{tabular}{l}
$\xi_u=-85.47+157.0 i$,\\
$\beta_u=29580 +5435 i$,\\
$\alpha_u= 6054$, $\gamma_u=167200$,\\
$|\xi_d|=14.53$, $|\beta_d|=442.5$,\\
 $\alpha_d= 2904$
\end{tabular}}
&\multirow{5}{*}{\begin{tabular}{c}
$\lambda_{1u}<0$\\
\\[0mm]
$\lambda_{2d}<0$
\end{tabular}}
&\qquad\qquad\multirow{5}{*}{
$\begin{matrix}
 -0.54 & 0.79 & 0.44 & -0.81 \\[2mm]
 0.98 & 0.13 & 0.43 & - \\[2mm]
 0.36 & 0.60 & 0.55 & - \\
\end{matrix}$
}
\\[11.0mm]
&&&&&
\\\cline{3-6} 
&&&&&\\[-2mm]
&\multirow{3}{*}{%
$M_{Id}=
\begin{pmatrix}
 0&|\xi_d|&0\\
|\xi_d|&0&|\beta_d|\\
0&|\beta_d|&\alpha_d
\end{pmatrix}$}&
\multirow{5}{*}{\bf b.}
&\multirow{3}{*}{\begin{tabular}{l}
$\xi_u=21.04-284.5 i$,\\
 $\beta_u= 18950+5890 i$,\\
 $\alpha_u= 1690$, $\gamma_u= 169000$,\\
$|\xi_d|=13.41$, $|\beta_d|= 392.6$,\\
$\alpha_d= 2857$
\end{tabular}}
&\multirow{5}{*}{\begin{tabular}{c}
$\lambda_{2u}<0$\\
\\[0mm]
$\lambda_{2d}<0$
\end{tabular}}&\qquad\qquad\multirow{5}{*}{$
\begin{matrix}
 -0.58 & -0.99 & -0.53 & -0.73 \\[2mm]
 -0.28 & 0.25 & -0.69 & - \\[2mm]
 0.68 & -0.26 & 0.000098 & - \\
\end{matrix}
$}
\\[10mm]
&&&&&
\\\hline
&&&&&\\[-2mm]
\multirow{9}{*}{\bf II}&
\multirow{4}{*}{$M_{IIu}=
\begin{pmatrix}
 0&0&|\xi_u|\\
0&\alpha_u&|\beta_u|\\
|\xi_u|&|\beta_u|&\gamma_u
\end{pmatrix}$}&\multirow{4}{*}{\bf a.}&\multirow{4}{*}{\begin{tabular}{l}
$|\xi_u|=431.5$, $|\beta_u|= 7251$,\\
 $\alpha_u= 957.9$, $\gamma_u= 171200$,\\
$\xi_d=4.316 + 14.26i$,\\
$\gamma_d=64.13$, $\alpha_d= 2969$
\end{tabular}}&\multirow{4}{*}{\begin{tabular}{c}
$\lambda_{1u}<0$\\
\\[0mm]
$\lambda_{1d}<0$
\end{tabular}}
&\qquad\qquad\quad\multirow{4}{*}{$
\begin{matrix}
 0.12 & 0.86 & 0.37 & 0.89 \\[1mm]
 0.52 & 0.51 & -0.47 & - \\[1mm]
 0.98 & 0.49 & 0.53 & -\\
\end{matrix}
$}\\[7mm]
&&&&&\\\cline{3-6} 
&&&&&\\[-2mm]
&\multirow{2}{*}{$M_{IId}=
\begin{pmatrix}
 0&\xi_d&0\\
\xi_d^*&\gamma_d&0\\
0&0&\alpha_d
\end{pmatrix}$}&\multirow{5}{*}{\bf b.}&\multirow{4}{*}{\begin{tabular}{l}
$|\xi_u|=426.3$, $|\beta_u|=7336$,\\
 $\alpha_u= 868.1$, $\gamma_u=172500$,\\
$\xi_d=-4.152-13.81i$,\\
$\gamma_d=-62.50$, $\alpha_d=2916$
\end{tabular}}&
\multirow{4}{*}{\begin{tabular}{c}
$\lambda_{1u}<0$\\
\\[0mm]
$\lambda_{2d}<0$
\end{tabular}}
&\qquad\qquad\quad\multirow{4}{*}{$
\begin{matrix}
 0.55 & 0.81 & 0.85 & 0.96 \\[1mm]
 0.62 & -0.97 & 0.63 & - \\[1mm]
 0.72 & 0.38 & 0.052 & - \\
\end{matrix}
$}\\[7mm]
&
&&&&\\\hline
\end{tabular}
\captionof{table}{
Patterns for quark mass matrices with five
texture zeros. The Wolfenstein parameters  for the CKM mixing matrix and the quark masses are reproduced with deviations
below $ 1\sigma $ level. In the last column $P_{A}=\frac{A_{\text{WB}}-A_{\text{PDG}}}{\Delta A}$, where $A_{\text{WB}}$
 and $A_{\text{PDG}}$ are the values for $A$ from the WB transformation and the PDG best fit, respectively.  $\Delta A$
is  the uncertainty for $A$ reported in the PDG.}
\label{t2}
}
\end{center}
\end{widetext}

\section{Mass matrices with five texture zeros}
\label{sIV}
As it is well known in the literature, for a given texture it is possible to establish relations
between the quark masses, the mixing angles and the CP violation phase of the CKM matrix, so that,  a study of these relations is important to shed light on  
the underlying symmetries of the flavor physics.
The five-zero textures for the quark mass matrices given in Table~\ref{t2} are viable models according to the latest data for the current quark masses and the CKM mixing matrix  parameters  at the $Z$ scale. 
In what follows we will consider various cases to implement quark mass matrices with five texture-zeros.
\subsection{Case I}
In this configuration,  the down-type quark mass matrix  contains three texture zeros, two of them on the diagonal, corresponding to the case~I in Table~\ref {t2}, which has the following analytical structure for quark mass matrices 
{
\begin{equation}
\label{5.1y}
\begin{split}
M_{Iu}&=
P^\dag\begin{pmatrix}
 0&0&|\xi_u|\\
0&\alpha_u&|\beta_u|\\
|\xi_u|&|\beta_u|&\gamma_u
\end{pmatrix}P,
\\
M_{Id}&=
\begin{pmatrix}
 0&|\xi_d|&0\\
|\xi_d|&0&|\beta_d|\\
0&|\beta_d|&\alpha_d
\end{pmatrix}, 
\end{split}
\end{equation}}
where all the phases are reduced to those contained in the diagonal matrix
$P=\textrm{diag}(e^{-i\phi_{\xi_u}},e^{-i\phi_{\beta_u}},1)$
(with  
$\phi_{\beta_u}\equiv 
\arg(\beta_u)$ and $\phi_{\xi_u}\equiv \arg(\xi_u)$)
which comes from doing a WB transformation, in such a way that the phases of $M_{Id}$ are absorbed in $P$. So we have 7 real parameters and 2 phases, to reproduce 10 physical quantities: 6  quark masses,  3 mixing angles and the  CP violating phase of the CKM mixing matrix, which implies that relations between masses and mixing angles can be established in the quark sector.
The five-zero texture deduced in~\eqref{5.1y} is not a Fritzsch texture of those studied in~\cite{Fritzsch:1999ee}.
Even though they are not identical, the mass matrices~\eqref{5.1y} can be diagonalized with the help of the matrix~\eqref{32x}. Let's use the permutation matrix
{\footnotesize $P_2=[(1,0,0),(0,0,1),(0,1,0)]$}, 
to bring the  up-type quark  mass matrix  to the form
 {\footnotesize$ M_u
=P^\dag P_2\begin{pmatrix}
 0&|\xi_u|&0\\
|\xi_u|&\gamma_u &|\beta_u|\\
0&|\beta_u|&\alpha_u
\end{pmatrix}P_2 P$},
in such a way that the internal matrix corresponds to that in~\eqref{30}. 
Therefore, the diagonalization matrix is the unitary
matrix $P^\dag \: P_2 \: U_u$, where $ U_u $ is defined in~\eqref{32x}, for the case $q=u$.  According to~\eqref{3.18y} the other mass matrix in~\eqref{5.1y}, $ M_ {Id} $, can be diagonalized if we make $ \alpha_d = \lambda_ {1d} + \lambda_ {2d} + \lambda_ {3d}$.

\begin{widetext}
From~\eqref{e3.4} the mass matrix parameters are:

\begin{subequations}
\label{e3.4x}
\begin{align}
\label{3.18}
 \gamma_u&=\mp m_u\pm m_c+m_t-\alpha_u,\\
\label{34a}
|\beta_u|&=\sqrt{\frac{(\alpha_u\pm m_u)(\alpha_u\mp 
m_c)(m_t-\alpha_u)}{\alpha_u}},\\
\label{35a}
|\xi_u|&=\sqrt{\frac{m_u\,m_c\,m_t}{\alpha_u}},
\\
 \alpha_d&=m_d-m_s+m_b,\\
\label{34az}
|\beta_d|&=\sqrt{\frac{(m_b-m_s)(m_d+m_b)(m_s-m_d)}{m_d-m_s+m_b}},\\
\label{35az}
|\xi_d|&=\sqrt{\frac{m_d\,m_s\,m_b}{m_d-m_s+m_b}},
\end{align}
\end{subequations}
where for  the eigenvalues of $M_{Iu}$  we have considered two possible cases $\lambda_{1u}<0$~(upper sign) and $\lambda_ {2u} <0 $~(lower sign). $ \alpha_u $ is a free parameter which, 
according to the equations~\eqref{e3.5s}, takes values in the intervals:
\begin{subequations}
\label{e4.3}
\begin{align}
\label{e4.3x}
 m_c\le \alpha_u\le m_t&\quad\text{for}\quad \lambda_{1u}<0,\\
 m_u\le \alpha_u\le m_t&\quad\text{for}\quad \lambda_{2u}<0.
\end{align}
\end{subequations}
The diagonalization matrices for $M_{Iu}$ and $M_{Id}$ in~\eqref{5.1y} are
{\small
\begin{equation}
 U_{Iu}=
\begin{pmatrix}
e^{i (\phi_{\xi_u}+ \theta_{1u})} \sqrt{\frac{m_c m_t (\alpha_u\pm 
m_u)}{\alpha_u (m_c+m_u)
(m_t\pm m_u)}} & \pm e^{i (\phi_{\xi_u}+ \theta_{2u})} \sqrt{\frac{(\alpha_u\mp 
m_c) m_t m_u}{\alpha_u (m_t\mp m_c) (m_c+m_u)}} & e^{i (\phi_{\xi_u}+ 
\theta_{3u})} \sqrt{\frac{m_c (m_t-\alpha_u) m_u}{\alpha_u (m_t\mp m_c) (m_t\pm 
m_u)}} \\
 \pm e^{i( \phi_{\beta_u}+ \theta_{1u})} \sqrt{\frac{(\alpha_u\mp m_c) 
(m_t-\alpha_u) m_u}
{\alpha_u (m_c+m_u) (m_t\pm m_u)}} & -e^{i (\phi_{\beta_u}+ \theta_{2u})} 
\sqrt{\frac{m_c (m_t-\alpha_u) (\alpha_u\pm m_u)}{\alpha_u (m_t\mp m_c) 
(m_c+m_u)}} &
e^{i (\phi_{\beta_u}+ \theta_{3u})} \sqrt{\frac{(\alpha_u\mp m_c) 
m_t (\alpha_u\pm m_u)}{\alpha_u (m_t\mp m_c) (m_t\pm m_u)}} \\
 \mp e^{i \theta_{1u}} \sqrt{\frac{m_u (\alpha_u\pm m_u)}{(m_c+m_u) (m_t\pm 
m_u)}} & e^{i \theta_{2u}} 
\sqrt{\frac{m_c (\alpha_u\mp m_c)}{(m_t\mp m_c) (m_c+m_u)}} & e^{i 
\theta_{3u}}\sqrt{\frac{m_t (m_t-\alpha_u)}{(m_t\mp m_c) (m_t\pm m_u)}} \\
\end{pmatrix},
\end{equation}}
{\small
\begin{equation}
 U_{Id}=
\begin{pmatrix}
 e^{i \theta_{1d}} \sqrt{\frac{m_b (m_b-m_s) m_s}{(m_b-m_d) (m_d+m_s) 
(m_b+m_d-m_s)}} &
-e^{i \theta_{2d}} \sqrt{\frac{m_b (m_b+m_d) m_d}{(m_d+m_s) (m_b+m_d-m_s) 
(m_b+m_s)}} & \sqrt{\frac{m_d (m_s-m_d) m_s}{(m_b-m_d) (m_b+m_d-m_s) 
(m_b+m_s)}} 
\\
 e^{i \theta_{1d}} \sqrt{\frac{m_d (m_b-m_s)}{(m_b-m_d) (m_d+m_s)}} & e^{i 
\theta_{2d}}
\sqrt{\frac{(m_b+m_d) m_s}{(m_d+m_s) (m_b+m_s)}} & \sqrt{\frac{m_b 
(m_s-m_d)}{(m_b-m_d) (m_b+m_s)}} \\
 -e^{i \theta_{1d}} \sqrt{\frac{m_d (m_b+m_d) (m_s-m_d)}{(m_b-m_d) (m_d+m_s) 
(m_b+m_d-m_s)}} 
& -e^{i \theta_{2d}} \sqrt{\frac{(m_b-m_s) m_s (m_s-m_d)}{(m_d+m_s) 
(m_b+m_d-m_s) (m_b+m_s)}} & \sqrt{\frac{m_b (m_b+m_d) (m_b-m_s)}{(m_b-m_d) 
(m_b+m_d-m_s) (m_b+m_s)}} \\
\end{pmatrix}
\end{equation}}
\end{widetext}
where the non-physical phases
$\theta_{1u},\theta_{2u},\theta_{3u},\theta_{1d}$ and
$\theta_{2d}$ 
 are necessary in order to adjust
our theoretical prediction for the CKM to the established convention. 
%
\begin{table}[htbp]
\begin{tabular}{c|cc|cc}
&\multicolumn{2}{ c}{\bf Case I}&\multicolumn{2}{ c}{\bf Case II}\\\hline
&$\begin{matrix}
 \lambda_{1u}<0\\
 \lambda_{2d}<0
\end{matrix}$&$\begin{matrix}
 \lambda_{2u}<0 \\
 \lambda_{2d}<0
\end{matrix}$&$\begin{matrix}
 \lambda_{1u}<0\\
 \lambda_{1d}<0
\end{matrix}$&$\begin{matrix}
 \lambda_{1u}<0\\
 \lambda_{2d}<0
\end{matrix}$\\\hline
$\theta_{1u}$&$-1.423$&$-2.844$&$-1.975$&$-1.991$ \\
$\theta_{2u}$&$0.6701$&$1.856$&$0$&$0$ \\
$\theta_{3u}$&$-0.004737$&$-0.004617$&$0$&$0$ \\
$\theta_{1d}$&$0.6360$&$1.930$&$3.025$&$-0.1351$ \\
$\theta_{2d}$&$-2.285$&$-0.9766$&$3.148$&$3.148$\\
$ \phi_{\xi_u}$ &$2.069$&$-1.497$&-&-\\
$ \phi_{\beta_u}$ &$0.1817$&$0.3015$&-&-\\
$ \phi_{\xi_d}$&-&-&$1.277$&$-1.863$\\
$\alpha_u $~(MeV)&$ 6054$&$1690$&$957.9$&$868.1$   \\
$m_u$~(MeV)& $1.792$&$1.268$& $1.599$&$1.642$\\
$m_c$~(MeV)& $625.5$&$633.2$& $650.2$&$555.7$\\
$m_t$~(MeV)& $172600$&$171300$& $171500$&$172900$\\ 
$m_d$~(MeV)&$2.993$&$3.148$&$3.292$&$3.166$\\
$m_s$~(MeV)&$68.93$&$56.12$&$67.42$&$65.66$\\
$m_b$~(MeV)&$2970$&$2910$&$2969$&$2916$\\\hline
\end{tabular}
\caption{Fit parameters.}
\label{num1}
\end{table}
%
To obtain the leading order~(LO) terms that contribute  to the CKM mixing matrix $V= U_{Iu}^\dag U_{Id}$ we use the hierarchy of the quark masses~\eqref{19d}. The analytical results for the LO CKM  entries are summarized in Table~\ref{t3}. There are several aspects to highlight about the case~I:
\begin{itemize}
 \item 
In the SM the inputs $|V_{cs}|\approx|V_{tb}| \approx 1$ then the free parameter must satisfy $ \alpha_u \ll m_t$,
hence  $\alpha_u/m_t \ll 1 $. Also, due to the condition~\eqref{e4.3x}, we have $\alpha_u \gg m_u$. 
\item
The free parameter $\alpha_u/m_t$ is only relevant for
the real parts of the matrix elements $V_{tb}$~(although $\alpha_u/m_t\ll1$   this matrix element is very precisely determined) and $V_{ub}$. 
For the matrix elements  $V_{ts}$, $V_{cb}$, $V_{ub}$  and $V_{td}$,
$\alpha_u/m_t$ is relevang for adjusting the CP violating phase. 
For the remaining matrix elements,  by neglecting linear  terms in $\alpha_u/m_t$ and $m_c/m_t$, the dominant contributions only depend on ratios between down-type quark masses.
\item
Relations can be established between the  elements of the mixing matrix whose LO terms only involve quark masses as shown in Table~\ref{t4}.
Some of these relations are well-known, for example the Gatto-Sartori-Tonin~(GST)~ (Eq.~2, in Table~\ref{t4})~\cite{b28}:
 $\tan\theta_{12}=|{V_{us}}/{V_{ud}}|=\sqrt{{m_d}/{m_s}}$,
 which is approximately fulfilled. 
Another important relation that we can find and that is  successfully verified, according to the experimental result~\eqref{eq1}, is given by the expression: $|V_{td}/V_{ts}|\approx\sqrt{m_d/m_s}$~\cite{Fritzsch:2021lzb}. On the other hand, our analysis also allows us to stablish that the relation $|V_{ub}/V_{cb}|$ does not coincide with the result $\sqrt{m_u/m_c}$, in accordance with the experimental data~\eqref{eq1}.
\item
The best fit for the mass matrix parameters~\eqref{5.1y} are shown in Table~\ref{num1}.
\end{itemize}

\begin{table}[htbp]
\begin{tabular}{c|cc|c}
\multicolumn{2}{c}{Relations}&{\bf Case I}&{\bf Case II}\\\hline
1&$\left|\frac{V_{cs}}{V_{ud}\,V_{tb}}\right|$&$1+\cdots$&$1+\cdots$\\
2&$\left|\frac{V_{us}}{V_{ud}}\right|$&$\sqrt{\frac{m_d}{m_s}}+\cdots$&$\sqrt{\frac{m_d}{m_s}}+\cdots$\\
3&$\left|\frac{V_{cd}}{V_{ud}\,V_{tb}}\right|$&$\sqrt{\frac{m_d}{m_s}}\,+\cdots$&$\sqrt{\frac{m_d}{m_s}}+\cdots$\\
4&$\left|\frac{V_{ts}}{V_{ud}\,V_{cb}}\right|$&$1+\cdots$&$1+\cdots$\\
5&$\left|\frac{V_{td}}{V_{ud}\,V_{cb}}\right|$&$\sqrt{\frac{m_d}{m_s}}+\cdots$&-\\
6&$\left|\frac{V_{cs}}{V_{tb}}\right|$&$\sqrt{\frac{m_s}{m_s+m_d}}
+\cdots$&$\sqrt{\frac{m_s}{m_s+m_d}}
+\cdots$\\
7&$\left|\frac{V_{cs}}{V_{us}\,V_{tb}}\right|$&$\sqrt{\frac{m_s}{m_d}}
+\cdots$&$\sqrt{\frac{m_s}{m_d}}
+\cdots$\\
8&$\left|\frac{V_{cs}}{V_{cd}}\right|$&$\sqrt{\frac{m_s}{m_d}}+\cdots$&$\sqrt{\frac{m_s}{m_d}}+\cdots$\\
9&$\left|\frac{V_{ts}\,V_{tb}}{V_{cs}\,V_{cb}}\right|$&$1+\cdots$&$1+\cdots$\\
10&$\left|\frac{V_{td}\,V_{tb}}{V_{cs}\,V_{cb}}\right|$&$\sqrt{\frac{m_d}{m_s}} 
+\cdots$&-
\\
11&$\left|\frac{V_{cd}}{V_{tb}}\right|$&$\sqrt{\frac{m_d}{m_s+m_d}}
+\cdots$&$\sqrt{\frac{m_d}{m_s+m_d}}
+\cdots$\\
12&$\left|\frac{V_{cd}}{V_{us}\,V_{tb}}\right|$&$1+\cdots$&$1+\cdots$\\
13&$\left|\frac{V_{ts}}{V_{us}\,V_{cb}}\right|$&$\sqrt{\frac{m_s}{m_d}}+\cdots$&$\sqrt{\frac{m_s}{m_d}}+\cdots$\\
14&$\left|\frac{V_{td}}{V_{us}\,V_{cb}}\right|$&$1+\cdots$&-
\\
15&$\left|\frac{V_{ts}\,V_{tb}}{V_{cd}\,V_{cb}}\right|$&$\sqrt{\frac{m_s}{m_d}}
+\cdots$&$\sqrt{\frac{m_s}{m_d}}
+\cdots$\\
16&$\left|\frac{V_{td}\,V_{tb}}{V_{cd}\,V_{cb}}\right|$&$1+\cdots$&- \\
17&$\left|\frac{V_{ts}}{V_{cb}}\right|$&$\sqrt{\frac{m_s}{m_s+m_d}}+\cdots$&$\sqrt{\frac{m_s}{m_s+m_d}}+\cdots$\\
18&$\left|\frac{V_{td}}{V_{ts}}\right|$&$\sqrt{\frac{m_d}{m_s}}+\cdots$&-\\
19&$\left|\frac{V_{td}}{V_{cb}}\right|$&$\sqrt{\frac{m_d}{m_s+m_d}}+\cdots$&-
\end{tabular}
\caption{Leading order relations between the CKM matrix elements.}
\label{t4}
\end{table}

\subsection{Case II}
Another viable analytical texture in Table~\ref {t2} is the
case~II, with quark mass matrices given by
{
\begin{equation}
\label{5.1xx}
\begin{split}
M_{IIu}&=
\begin{pmatrix}
 0&0&|\xi_u|\\
0&\alpha_u&|\beta_u|\\
|\xi_u|&|\beta_u|&\gamma_u
\end{pmatrix},
\\
M_{IId}&=
\begin{pmatrix}
 0&|\xi_d|\,e^{i\phi_{\xi_d}}&0\\
|\xi_d|\,e^{-i\phi_{\xi_d}}&\gamma_d&0\\
0&0&\alpha_d
\end{pmatrix}.
\end{split}
\end{equation}}
In this case we have only one phase,
$ \phi_ {\xi_d}$, responsible for the CP violation. And there are 7 real parameters. This texture is a Fritzsch-type~\cite{Fritzsch:1999ee}.

As in the previous case we can obtain relations between
the elements of the CKM and the quark masses.
The structure of the matrix  $M_{IIu}$ is similar to the one given in  Eq.~\eqref{5.1y}, without including phases, so it can be inferred that the diagonalization matrix for this case is:
$P_2\:U_u$  with
$P_2=[(1,0,0),(0,0,1),(0,1,0)]$ and $U_u$ defined in~\eqref{32x} for $q=u$.

The matrix $M_{IId}$ in~\eqref{5.1xx} has a zero structure like the one given in~\eqref{30} 
with $|\beta_{q}|=0$, so that, there are several possibilities to be considered:
$\alpha_d=\lambda_{1d}>0$, $|\xi_d|=\sqrt{-\lambda_{2d}\,\lambda_{3d}}$  and 
$\gamma_d=\lambda_{2d}+\lambda_{3d}$; or $\alpha_d=\lambda_{2d}>0$, 
$|\xi_d|=\sqrt{-\lambda_{1d}\,\lambda_{3d}}$ and
$\gamma_d=\lambda_{1d}+\lambda_{3d}$; or $\alpha_d=\lambda_{3d}>0$, 
$|\xi_d|=\sqrt{-\lambda_{1d}\,\lambda_{2d}}$ and
$\gamma_d=\lambda_{1d}+\lambda_{2d}$~\cite{Giraldo:2011ya}. 
From the last option we obtain the two cases with the best agreement with the data, as reported in Table~\ref{t2}.
Here the diagonalization matrix for $M_{IId}$ is $P^\dag_d\,U_d$ with $U_d$ as given in~\eqref{32x} for $q=d$ and  $P_d=\textrm{diag}(e^{-i\phi_{\xi_d}},1,1)$. 
The parameters of the mass matrices~\eqref{5.1xx}, according to the relations~\eqref{e3.4} are given by:
\begin{subequations}
\label{e3.4xx}
\begin{align}
\label{3.18x}
 \gamma_u&=- m_u+ m_c+m_t-\alpha_u,\\
\label{34ax}
|\beta_u|&=\sqrt{\frac{(\alpha_u+ m_u)(\alpha_u- m_c)(m_t-\alpha_u)}{\alpha_u}},\\
\label{35ax}
|\xi_u|&=\sqrt{\frac{m_u\,m_c\,m_t}{\alpha_u}},
\\
 \alpha_d&=m_b,\\
\label{34axz}
|\xi_d|&=\sqrt{m_d\,m_s},\\
\label{35axz}
\gamma_d&=\mp m_d\pm m_s,
\end{align}
\end{subequations}
%
%
where $\lambda_{1u} <0$; the upper sign, for $ \lambda_ {1d} <0$ and the lower sign, for $ \lambda_ {2d}<0$; 
and $ \alpha_u>0$ is a free parameter   in the range:
\begin{equation}
 m_c\le \alpha_u\le m_t.
 \label{e3.8}
\end{equation}
In this case, the diagonalization matrices of the mass operators~\eqref{5.1xx}, are:
\begin{widetext}
{
\begin{equation}
 U_{IIu}=
\begin{pmatrix}
e^{i \theta_{1u}} \sqrt{\frac{m_c m_t (\alpha_u+ 
m_u)}{\alpha_u (m_c+m_u)
(m_t+ m_u)}} &  e^{i \theta_{2u}} \sqrt{\frac{(\alpha_u- 
m_c) m_t m_u}{\alpha_u (m_t- m_c) (m_c+m_u)}} & e^{i  
\theta_{3u}} \sqrt{\frac{m_c (m_t-\alpha_u) m_u}{\alpha_u (m_t- m_c) (m_t+ 
m_u)}} \\
  e^{i\theta_{1u}} \sqrt{\frac{(\alpha_u- m_c) 
(m_t-\alpha_u) m_u}
{\alpha_u (m_c+m_u) (m_t+ m_u)}} & -e^{i \theta_{2u}} 
\sqrt{\frac{m_c (m_t-\alpha_u) (\alpha_u+ m_u)}{\alpha_u (m_t- m_c) 
(m_c+m_u)}} &
e^{i  \theta_{3u}} \sqrt{\frac{(\alpha_u- m_c) 
m_t (\alpha_u+ m_u)}{\alpha_u (m_t- m_c) (m_t+ m_u)}} \\
 - e^{i\theta_{1u}} \sqrt{\frac{m_u (\alpha_u+ m_u)}{(m_c+m_u) (m_t+ 
m_u)}} & e^{i \theta_{2u}} 
\sqrt{\frac{m_c (\alpha_u- m_c)}{(m_t- m_c) (m_c+m_u)}} & e^{i 
\theta_{3u}}\sqrt{\frac{m_t (m_t-\alpha_u)}{(m_t- m_c) (m_t+ m_u)}} \\
\end{pmatrix},
\end{equation}}

{\small
\begin{equation}
 U_{IId}=
\begin{pmatrix}
 e^{i (\phi_{\xi_d}+\theta_{1d})} \sqrt{\frac{m_s}{m_d+m_s}} & \pm e^{i (\phi_{\xi_d}+\theta_{2d})} \sqrt{\frac{m_d}{m_d+m_s}} & 0 \\
 \mp e^{i \theta_{1d}} \sqrt{\frac{m_d}{m_d+m_s}} & e^{i \theta_{2d}} \sqrt{\frac{m_s}{m_d+m_s}} & 0 \\
 0 & 0 & 1 \\
\end{pmatrix}.
\end{equation}}
\end{widetext}
The best fit parameters are shown in Table~\ref{num1}.

Taking into account the hierarchy of the quark masses, Eq.~\eqref{19d},
and the interval for the parameter $\alpha_u$, Eq.~\eqref{e3.8},
to LO the entries of the CKM, $V = U_u^\dag U_d$, are summarized in Table~\ref{t3},~case~II; from these results we  conclude that:
\begin{itemize}
\item 
As in the case~I,  $ | V_ {cs} | \approx |V_{tb}| \approx1 $,
such that $\alpha_u \ll m_t$. Also, due to~\eqref{e3.8},  $\alpha_u \gg m_u$. 
Again we verify that the relation $|V_{ub}/V_{cb}|\neq\sqrt{m_u/m_c} $ is not satisfied. Although the relationship $|V_{td}/V_{ts}|\approx\sqrt{m_d/m_s}$ is not verified in principle for this case; we can see from Table~\ref{t3}, for the case~II, that if we omit the second term of the approximation for $V_{td}$, it is enough to reproduce this relationship. As it is shown numerically, for this case high order contributions are neglegible.
 \item 
The CKM matrix  elements, $V_{ts}$, $V_ {cb}$, $V_{ub}$, $V_{td}$  depend heavily on the $\alpha_u$ parameter, the remaining elements depend on ratios between  down quark masses.
Only the $V_{td}$ matrix element has information about the phase, which in turn depends on the ratio $\alpha_u/m_t$, 
which is a noticeable difference with respect to case I. 
\item 
As in the case~I, the LO relations  between the  CKM elements  involving only quark masses are shown in Table~\ref{t4}.
\item
Although, the results are similar to the expressions given in Table~\ref{t4} for the case I, 
for the case II the relations~5, 10, 14, 16, 18 and 19 are absent~(the corresponding expressions are cumbersome).
\end{itemize}
%
\section{Conclusions}
\label{sV}
Using the  WB transformation method~\cite{Giraldo:2011ya,Giraldo:2015ffa}, we found  configurations for the quark mass matrices with the maximum number of possible texture zeros. To accomplish this,  we start from the general basis~\eqref{3} and~\eqref{4}, from which the expressions~\eqref{2.9x} and~\eqref{2.10} can be obtained, respectively. 
Modulo permutations, only the configurations shown in Table~\ref{ta1}, for mass matrices with one or two zeros in the 
diagonal, are possible.  
From these patterns we obtained the cases I and II in Table~\ref{t2},
 corresponding to the  five-zero textures in Eq.~\eqref{5.1y} and Eq.~\eqref{5.1xx}, respectively,
which reproduce the quark masses, mixing angles and the  CP violation phase, with  deviations from the experimental values below  1$\sigma$ level.

The first case has nine free parameters: 7 real and 2 phases, while the second case has eight free parameters: 7 real and 1 phase. In both cases, it is necessary to reproduce ten physical quantities: 6 quark masses, 3 mixing angles and the CP violation phase, the lack of balance between the number of free parameters  and the physical quantities implies  physical relations between the quark  masses and the CKM mixing angles, which are reported   in Table~\ref{t3}.
Additionally, the relation GST~\cite{b28} is maintained and we can adjust the CP violation phase of the SM. 
Additionally, our five-zero texture models reproduce the experimental quantities~\eqref{eq1} deviating from the experimental central value by at most  $1\, \sigma $ level. First, for both cases~I and~II, we can verify that the relation $|V_{td}/V_{ts}|\approx\sqrt{m_d/m_s}$ is satisfied, while the relation $|V_ {ub}/V_{cb}|\neq\sqrt{m_u /m_c}$ is not,  as already indicated in previous works~\cite{Tanabashi:2018oca, UTfit, CKMfitter}. In some cases, even with the LO approximations given in Table~\ref{t3} we can reproduce the results~\eqref{eq1}. We have several free parameters to adjust the physical quantities: 
the CP-violating phases,  the calibration phases $\theta_{q i}$ with $q=u,d$, in the diagonalization matrices, and  the real parameter~$\alpha_u$. In our analysis the analytical LO expressions  for the case~I, with $ \lambda_ {1u} <0 $ and for the case~II, with $ \lambda_ {1d} <0 $, are enough to keep all the observables inside of the error bars. For the other cases, $\lambda_{2u}<0$ and $\lambda_ {2d}<0$, satisfactory results were not achieved with the LO approximations  provided in Table~\ref{t3}. In these cases, the complete expressions must be taken into account in order to get a good fit.

The  case~I is an original proposal which was not considered in the  Fritzsch original work~\cite{Fritzsch:1999ee}  nor in  later studies.
Case II has been widely considered in the literature~\cite{Fritzsch:1999ee,Verma:2017ppl,Ludl:2015lta,Mahajan:2009wd,Ponce:2013nsa,Desai:2000bu}, 
but in our approach, we take a negative eigenvalue (which has not been considered previously) for the mass of the lightest down quark, that is,
$\lambda_ {1d}<0$. Here, it should be mentioned that, without losing generality, only one negative eigenvalue is necessary for each mass matrix~\cite{Giraldo:2011ya,Giraldo:2015ffa}. Also, it is important to say that the relations in Table~\ref{t3} are comparable to the results reported in~\cite{b4,b7,Verma:2017ppl,Kaundal:2019njx,Ponce:2013nsa}.

The purpose of the texture zeros for quark mass matrices is to find relations between quark masses and  the flavor mixing parameters in  consistency  with the experimental data~\cite{Fritzsch:1999ee}. For the  textures deduced in this work,   the quark mass ratios contribute significantly to  the flavor mixing parameters as shown in Table~\ref {t4}; In Table~\ref {t3}, it is possible to observe additional contributions (not exclusively dependent on the quark masses) which also depend on the free parameter $\alpha_\mu$ and on the phases responsible for the CP violation.
It is important to highlight that the LO contributions to the relations involving the CKM matrix elements mainly depend on ratios of down-type quark masses. 
The relations reported in this manuscript, could be useful to disentangle the underlying symmetries under the mass scales in the SM.


\section*{Acknowledgments}
In these subjects we acknowledge the collaboration of W. A. Ponce, R. Benavides and L. Muñoz.
We also acknowledge financial support from VIIS in the Universidad de Nariño, Approval Contract No. 024 and 160, project numbers: 1048, 1928  and 2172. 
\appendix

\section{Quark mass matrices and the CKM mixing matrix}
\label{sII}
The parameters of the CKM  are reported at the $Z$ pole scale  $\mu= M_Z $, hence the same scale is used  to evaluate  the current quark masses~(in MeV)~\cite{Xing:2011aa}., i.e.,
{
{\small
\begin{equation}
\label{17a}
\begin{split}
 m_u&=1.38^{+0.42}_{-0.41}\,,\: m_c=638^{+43}_{-84},\: m_t=172100\pm{1200}\,,\\
m_d&=2.82\pm0.48\,,\: m_s=57^{+18}_{-12}\,,\: m_b=2860^{+160}_{-60}.
\end{split}
\end{equation}}}
The  CKM unitary matrix~\cite{b19,b20,Tanabashi:2018oca} can be parameterized by three mixing angles and the CP violation  phase~\cite {b20}.  The form of this matrix in the standard parametrization is given by~\cite{b21}.
\begin{widetext}
{
\begin{equation}
\label{3.2}
 V=\begin{pmatrix}
    V_{ud}&V_{us}&V_{ub}
\\
V_{cd}&V_{cs}&V_{cb}
\\
V_{td}&V_{ts}&V_{tb}
   \end{pmatrix}=\begin{pmatrix}
    c_{12}\,c_{13}&s_{12}\,c_{13}&s_{13}\,e^{-i\delta}\\
-s_{12}\,c_{23}-c_{12}\,s_{23}\,s_{13}\,e^{i\delta}& 
c_{12}\,c_{23}-s_{12}\,s_{23}\,s_{13}\,e^{i\delta}&
s_{23}\,c_{13}\\
s_{12}\,s_{23}-c_{12}\,c_{23}\,s_{13}\,e^{i\delta}&-c_{12}\,s_{23}-s_{12}\,c_{23
}\,s_{13}\,e^{i\delta}&c_{23}\,c_{13}
   \end{pmatrix},
\end{equation}}
%
where $s_{ij}=\sin \theta_{ij}$,  $c_{ij}=\cos \theta_{ij}$,
the angles $\theta_{ij}$ are said to lie in the first quadrant, such that $\sin\theta_ {ij}, \cos\theta_ {ij}\ge0$. The
phase $\delta$ is responsible for all the CP violation phenomena in the flavor changing processes in SM.
For various applications it is useful to use the Wolfenstein parameterization~\cite{Tanabashi:2018oca}
{
\begin{equation}
\label{3.3}
 \begin{split}
  \lambda&= \sin\theta_{12},\quad\quad \quad\quad 
A=\frac{\sin\theta_{23}}{\sin^2\theta_{12}}, \quad
\quad
\rho=\frac{\sin\theta_{13}\cos\delta}{\sin\theta_{12}\sin\theta_{23}},\quad
\eta=\frac{\sin\theta_{13}\sin\delta}{\sin\theta_{12}\sin\theta_{23}}.
 \end{split}
\end{equation}}
The CKMfitter and UTfit Collaborations~\cite{CKMfitter,UTfit} provide updated  fits for the Wolfenstein parameters,
{
\begin{equation}
\label{3.4}
 \begin{split}
  \lambda&=0.22500_{-0.00100}^{+0.00100},\quad 
A=0.826^{+0.012}_{-0.012}, \quad
\rho=0.152^{+0.014}_{-0.014},\quad\eta=0.357_{-0.010}^{+0.010}.
 \end{split}
\end{equation}}
The best fit values for CKM matrix elements are
%
{\footnotesize
\begin{equation}
\label{2}
V=\begin{pmatrix}
 (0.97431\pm0.00012) & (0.22514\pm0.00055)& 
(0.00365\pm0.00010)e^{i(-66.8\pm2.0)^\circ} \\
 (-0.22500\pm0.00054)e^{i(0.0351\pm0.0010)^\circ} & 
(0.97344\pm0.00012)e^{i(-0.001880\pm0.000052)^\circ} & (0.04241\pm0.00065)
\\
(0.00869\pm0.00014)e^{i(-22.23\pm0.63)^\circ} & 
(-0.04124\pm0.00056)e^{i(1.056\pm0.032)^\circ} & (0.999112\pm0.000024)
\end{pmatrix}.
\end{equation}}
%
{\footnotesize
\begin{tabular}{c|p{5.3cm}|p{12cm}}
\hline
\hline
\multirow{2}{*}{\bf Case}&\multirow{2}{*}{\centerline{\bf Five-zero textures.}}
&\multirow{2}{*}{\centerline{\bf LO predictions for the CKM mixing matrix elements  $V_{\text{CKM}}$:}}
\\[+4mm]\hline
\multirow{25}{*}{I}&
{\begin{equation*}
\label{5.1}
\begin{split}
M_u&=
P^\dag\begin{pmatrix}
 0&0&|\xi_u|\\
0&\alpha_u&|\beta_u|\\
|\xi_u|&|\beta_u|&\gamma_u
\end{pmatrix}P,
\\
M_d&=
\begin{pmatrix}
 0&|\xi_d|&0\\
|\xi_d|&0&|\beta_d|\\
0&|\beta_d|&\alpha_d
\end{pmatrix}, 
\end{split}
\end{equation*} 
where {\small 
$P=\textrm{diag}(e^{-i\phi_{\xi_u}},e^{-i\phi_{\beta_u}},1)$}.

Besides $m_c<\alpha_u\ll m_t$.
\newline

With the upper sign~(``$-$'') for the case~(Ia), Table~\ref{t2}: \newline$ \lambda_{1u}<0$ and 
$\lambda_{2d}<0.$ 
\newline

With the lower sign~(``$+$'') for the case~(Ib), Table~\ref{t2}: \newline $ \lambda_{2u}<0$ and
$\lambda_{2d}<0.$}
&
{\begin{equation*}
\begin{split}
  |V_{ud}|&= \sqrt{\frac{m_s}{m_s+m_d}}+\cdots,
  \\
  |V_{cs}|&=\sqrt{\frac{m_s}{m_s+m_d} \left(1-\frac{ 
\alpha_u}{m_t}\right)}+\cdots,
  \\
  |V_{tb}|&=\sqrt{1-\frac{\alpha_u}{m_t}}+\cdots,
  \\
|V_{us}|&=\left|\sqrt{\frac{m_d}{m_s+m_d}}\,+\cdots\right|,
\\
|V_{cd}|&=\left|\sqrt{\frac{m_d}{m_s+m_d}\left(1-\frac{ 
\alpha_u}{m_t}\right)}\,+\cdots\right|,
\\
|V_{ts}|&= \left|\sqrt{\frac{
m_s}{ 
m_s+m_d}}
\left[
\sqrt{\frac{
m_s-m_d}{ 
m_b}\left(1-\frac{\alpha_u}{m_t}\right)}-e^{-i 
\phi_{\beta_u}}\sqrt{\frac{\alpha_u}{m_t}\mp\frac{m_c}{m_t}}\,
\right]+\cdots\right|,
\\
|V_{cb}|&= \left|\sqrt{\frac{m_s-m_d}{m_b 
}\left(1-\frac{\alpha_u}{m_t}\right)}-e^{i 
\phi_{\beta_u}}\sqrt{\frac{\alpha_u}{m_t}\mp\frac{m_c}{m_t}}\,
+\cdots\right|,
\\
|V_{ub}|&= 
\left|\sqrt{\frac{ m_u}{m_c}\frac{\alpha_u}{m_t}}-e^{-i 
\phi_{\beta_u}} \sqrt{\frac{ m_u(m_s-m_d)}{m_b 
}\left(\frac{1}{m_c}\mp\frac{1}{\alpha_u}\right)\left(1-\frac{\alpha_u}{m_t}
\right)}\right.
\\&\left.\mp e^{-i\phi_{\xi_u}} \sqrt{\frac{m_d\,m_s 
(m_s-m_d)}{m_b^3}}+\cdots\right|,
\\
|V_{td}|&= \left|\sqrt{\frac{m_d }{ 
m_s+m_d}}\left[
\sqrt{\frac{m_s-m_d }{ 
m_b}\left(1-\frac{\alpha_u}{m_t}\right)}- e^{-i 
\phi_{\beta_u}}\sqrt{\frac{\alpha_u}{m_t}\mp\frac{m_c
}{m_t}}\right]\,+\cdots\right|.
\end{split}
\end{equation*}}
\\\hline
\multirow{11}{*}{II}&{
\begin{equation*}
\label{5.1x}
\begin{split}
M_u&=
\begin{pmatrix}
 0&0&|\xi_u|\\
0&\alpha_u&|\beta_u|\\
|\xi_u|&|\beta_u|&\gamma_u
\end{pmatrix},
\\
M_d&=
\begin{pmatrix}
 0&|\xi_d|e^{i \phi_{\xi_d}}&0\\
|\xi_d|e^{-i\phi_{\xi_d}}&\gamma_d&0\\
0&0&\alpha_d
\end{pmatrix},
\end{split}
\end{equation*} 

 where $m_c<\alpha_u\ll m_t$, and\newline

Upper sign~($-$): for $ \lambda_{1u}<0$ and $\lambda_{1d}<0.$\newline

Lower sign~($+$): for $\lambda_{1u}<0$ and $\lambda_{2d}<0.$}&
{\begin{equation*}
\begin{split}
  |V_{ud}|&= \sqrt{\frac{m_s}{m_s+m_d}}+\cdots,
  \\
  |V_{cs}|&=\sqrt{\frac{m_s 
}{m_s+m_d}\left(1-\frac{\alpha_u}{m_t}\right)}\,+\cdots,
  \\
  |V_{tb}|&=\sqrt{1-\frac{\alpha_u}{m_t}}\,+\cdots,
  \\
|V_{us}|&=\left|\sqrt{\frac{m_d}{m_s+m_d}}\,+\cdots\right|,
\\
|V_{cd}|&=\left|\sqrt{\frac{m_d 
}{m_s+m_d}\left(1-\frac{\alpha_u}{m_t}
\right)}\,+\cdots\right|,
\\
|V_{ts}|&= \left|\sqrt{\frac{ m_s 
}{m_s+m_d}\left(\frac{\alpha_u}{m_t}
-\frac{m_c}{m_t}\right)}\,+\cdots\right|,
\\
|V_{cb}|&= 
\left|\sqrt{\frac{\alpha_u}{m_t}-\frac{m_c}{m_t}}\,+\cdots\right|,
\\
|V_{ub}|&= \left|\sqrt{\frac{m_u}{m_c}\frac{\alpha_u 
}{m_t}}\,+\cdots\right|,
\\
|V_{td}|&= \left|\sqrt{\frac{ m_d 
}{m_s+m_d}\left(\frac{\alpha_u}{m_t}
-\frac{m_c}{m_t}\right)}\mp e^{i \phi_{\xi_d}}\sqrt{\frac{ 
m_s\,m_c \,
m_u}{m_s+m_d}\frac{1}{m_t}\left(\frac{1}{\alpha_u 
}-\frac{1}{m_t}\right)}\,+\cdots\right|.
\end{split}
\end{equation*}}
\\\hline
\end{tabular}
}
\captionof{table}{Cases I and II for the quark mass matrices with five texture-zeros. And their corresponding LO predictions for the CKM elements.}
\label{t3}
\end{widetext}

\bibliographystyle{abbrv}
\bibliography{textura}
\end{document}